\documentclass[11pt,twoside]{article}

\usepackage{asp2004}
\usepackage{epsf}
\usepackage{lscape}
\usepackage{amsmath,amssymb}
\usepackage{graphicx}

\begin{document}
\title{Radiatively Driven Winds of OB Stars -- from Micro to Macro}
\author{Ji\v{r}\'{i} Krti\v{c}ka}   %%% Fill in author names
\affil{%
%Kr4: Masarykova univerzita,
\'Ustav teoretick\'e fyziky a astrofyziky P\v{r}F MU, Kotl\'a\v rsk\'a 2,\\
%Ku4: Pak se div, ze Te nekdo pojmenuje Jiri Masaryk Krticka :-)
%	Nechces tu adresu zpresnit?
       CZ-611 37 Brno, Czech Republic, 
%Kr4: \\
%
       email: krticka@physics.muni.cz}    %%% Fill in author affiliations
\author{Ji\v{r}\'{\i} Kub\'at}
\affil{Astronomick\'y \'ustav
%Ku1:
AV \v{C}R,
CZ-251 65 Ond\v{r}ejov, Czech
	Republic, \\email: kubat@sunstel.asu.cas.cz}

\begin{abstract} 
We review basic physics of line-driven stellar winds of OB stars. We discuss
elementary processes due to which stellar winds are accelerated on a microscopic
level. We show how these microscopic processes may enable the outflow and how
they determine wind properties on a macroscopic level. 
%Kr1: Finally, 
We discuss
shortcomings of present wind theories and future wind model improvements.
\end{abstract}

%Kr1:
%

%%%%%%%%%%%%%%%%%%%%%%%%%%%%%%%%%%%%%%%%%%%%%%%%%%%

\section{Introduction}

Observations of many OB stars show that there
%Ku1: exist
exists
an outflow of
material from the stellar surface into the interstellar medium --
{\em
%Ku1:
the
stellar wind}.
The theoretical study of
%Ku1: hots star wind
hot star winds
started few years after the discovery that electromagnetic radiation
carries momentum that can be transferred to the matter 
%Kr2: during
in
the process of light scattering.
\citet{mil} and \citet{joel,joel2} studied the possibility of the
emission of high-speed atoms from stars.
\citet{milskvel} in a beautifully written paper showed the importance of
the Doppler effect for the line radiative acceleration.
However, for the next few decades the radiatively driven hot star winds
did
%Ku1: no attracted
not attract
much attention.
Modern studies of hot
%Ku1: stars
stars'
winds were stimulated mainly by UV observations of hot stars.
Pioneering
%Ku1: work
works
of \citet{lusol} and \citet*[hereafter CAK]{cak} serve as a basis for
present hot star wind theory. 

%Kr1:
Except a more general book by \citet{vetry},
%
%Kr1: there are many contemporary reviews of hot star wind physics. 
\citet{owoza} provides an elegant introduction to the
%Kr1: topic, 
hot star wind physics,
while \citet{kupul} or \citet{owopo}
can be consulted for a more detailed review. 
%Kr1: The recent review of \citet{lamta} discusses mostly the wind mass loss rates.
%Ku1: nevim, ale nemeli bychom zminit i ucebnici Lamers&Cassinelli?

%Kr1: prestehoval jsem
In this review we intend to answer two basic questions: 
First, what are the processes responsible for the acceleration of
the stellar wind of OB stars?
Second, what are the theoretical predictions of the wind structure?

%%%%%%%%%%%%%%%%%%%%%%%%%%%%%%%%%%%%%%%%%%%%%%%%%%%

\section{Stellar wind of hot stars: micro-view}

\newcommand{\zav}[1]{\left(#1\right)}

\begin{figure}
\begin{center}
\includegraphics{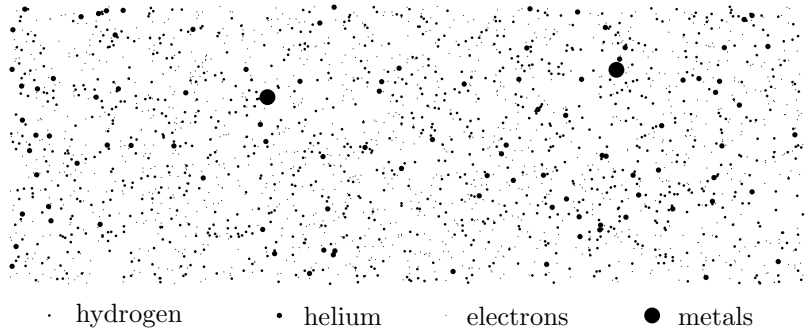}
\end{center}
\caption{A typical wind volume with 1000 H ions, 100 He ions,
%Ku1: 1200 electrons and 2 metallic ions}
%   nemas tam zahrnute elektrony od tech dvou iontu ...
2 metallic ions, and corresponding  number 
%Kr1:
(about 1200)
of electrons.}
\label{zvetsen}
\end{figure}

According to the theoretical studies of hot star winds, they are accelerated
due to the light scattering in lines of heavier elements, and
%Ku1:
to a lesser extent due to scattering
on free electrons.
%Ku1:
Hydrogen and helium are
%Kr1: inefficient absorbers.
mostly inefficient for wind driving.
From the microscopic view of stellar wind in Fig.~\ref{zvetsen} it is clear that
two
%Ku1: efficient
basic
processes are necessary to accelerate the bulk wind flow:
\begin{enumerate}
%Ku1:
\itemsep -1pt
\item process that transfers momentum from the
%Ku3: radiative
radiation
field to heavier ions
%Kr2: (and electrons),
and electrons,
\item process that transfers momentum from heavier ions 
%Kr2:
and electrons
to the bulk flow 
%Kr2: (H, He
(hydrogen and helium ions
-- mostly passive component).
\end{enumerate}

\subsection{How to transfer the
%Kr2: light momentum
radiation momentum
to heavier ions?}

To better understand transfer of the 
%Kr2: light momentum
radiation momentum
to heavier ions we
%Ku1: will study an
shall
%Kr1: describe
generally discuss
absorption and emission mechanisms.
The photon frequencies before absorption and after emission are $\nu$
and $\nu'$, respectively
%Kr1: presunuto
(note that the case $\nu^\prime=0$ means pure absorption).

In the general case $\nu\neq\nu'$
%Ku1: Tady je to cele nejake divne, piseme o rozptylu a popisujeme
%absorpci. Nevim, cos chtel presne rict, ale tak, jak to je, je to
%spatne. Misto \nu by tam melo byt \Delta\nu a system procesu, ktere
%ohrivaji nebo chladi je podstatne slozitejsi. Zkusil jsem to prepsat.
%Kr1: vracim a prepisuji
%Ku2: Porad je to nesrozumitelne. Musi se vse poradne vysvetlit, ale pak
%se to nevejde. Takhle napsane je to spatne. Nemenim to, ale nesouhlasim
%(as strongly as I can, jak by rekl Owocki).
the
%Ku3: energy transferred 
transferred energy
%
%Kr2: is about $\Delta E\approx h\nu$,
can be approximated as $\Delta E=h(\nu-\nu')\approx h\nu$,
%Ku2: I kdyz je to priblizne, je treba to vysvetlit, protoze kazdy by
%cekal $\Delta E\approx h\Delta\nu$
the
%Ku3: momentum transferred is 
transferred momentum
%
%Kr2:
then
about $\Delta p \approx h\nu/c$.
%Ku2: zase by tu melo byt \Delta\nu
From this we can estimate the change of the kinetic energy
%Kr2: of the atom
as
%Ku2: Nevim, jak muze foton predat energii atomu tak, ze se z ni stane
%kineticka energie toho atomu. Dokazu si ale predstavit, ze energie
%prispeje k tepelne tim, ze dojde ke srazce pri naslednem procesu,
%energii pak ale ziska jina castice, nejcasteji elektron.
$\Delta E_\text{kin}\approx\frac{1}{2}m_\text{i}
\zav{\Delta p /m_\text{i}}^2=\frac{1}{2}h^2\nu^2/\zav{m_\text{i}c^2}$. For a
typical UV radiation
%Ku3: however\linebreak
%
$\frac{1}{2}h^2\nu^2/\zav{m_\text{i}c^2}\ll \Delta E$,
and this means that most of
%Ku3: energy transferred
transferred energy
goes
%Kr2: to the internal energy (e.g.~to heating or cooling for $\nu<\nu'$)
finally to heating
%Ku3:
(for $\nu>\nu'$)
or cooling (for $\nu<\nu'$)
%Kr1: chaotic movement of metallic ions -- to {\em heating}
%(for $\nu<\nu'$ cooling)
of material
%Kr1:
and not to the
%Kr2:
macroscopic
kinetic energy.
This
explains why the most common interaction of material and light results in
heating or cooling.
%Kr1:
%either a part of incident photon energy is by means of collisional
%deexcitation transferred to the thermal energy (and results in heating)
%or by means of collisional excitation some energy is taken from the
%thermal pool (and results in cooling).
%All these processes are accompanied by momentum transfer.
%
%Ku1: However,
%Kr1:
On the contrary,
in the case $\nu\approx\nu'$
%Kr1:, the most important result of light scattering is
%Kr2: a significant part
most
of
%Ku3: energy transferred
transferred energy
goes to the
%Kr2:
macroscopic
kinetic energy
%Ku2: To zni divne, pokud je $\nu\approx\nu'$, tak kolik zbude na
%"significant part of energy"?
%Ku1:
%Kr1: only
%
%Kr1: the momentum transfer.
%Kr1: Consequently, 
and 
%Kr2: 
the irradiated 
material is accelerated (assuming 
%Kr1: that emission is isotropic).
isotropic emission).
%

%Kr2: Clearly, ideal
Most effective
processes for wind acceleration are those for which
$\nu\approx\nu'$.
%Ku1: Neni mi jasne proc jsou procesy s rozdilnou energii horsi pro
%urychlovani.
%Ku2: Porad mi to neni jasne.
This condition is fulfilled  for light scattering
%Kr1:
in lines and
on free electrons.
%Kr1: and for
%Ku1: absorption (scattering) - pro absorpci je \nu^\prime=0
%scattering
%
%of radiation in lines.
%
Both
%Kr1: these
processes are important for the acceleration of hot star winds.

\begin{figure}[ht]
\begin{center}
\resizebox{0.7\hsize}{!}{\includegraphics{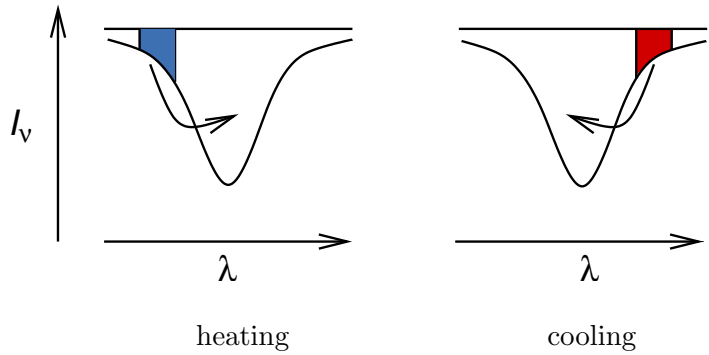}}
\end{center}
\hspace{4.5cm}heating\hspace{3.3cm} cooling
\caption{
%Ku1:
Schematic picture of
Doppler (Gayley-Owocki) heating/cooling.}
\label{doprof}
\end{figure}

However, even line scattering may become
%Ku1: inefficient.
less efficient.
This is basically connected with the fact that frequencies of absorbed
and emmited line radiation are slightly different.
This effect, introduced in the domain of hot star wind theory by
\citet{go} is called Doppler or Gayley-Owocki
heating/cooling.
An example of this effect is given in Fig.~\ref{doprof}.
Let us have an artificial light source that emits radiation only in a
very narrow wavelength interval corresponding only to the left (blue)
part of the line-profile (see Fig.~\ref{doprof}).
After the processes of absorption and emission the light from this
source is redistributed over all wavelengths of
%Ku1:
a
given line.
%Ku3: Clearly,
%
Some part of the radiative energy
%Kr2: is
has been
thermalized, emitted
radiation has lower energy, and the plasma can be heated by this
process.
On the other hand, let us
%Ku1: we
%
have a light source that emits radiation with wavelengths corresponding
only to the right (red) part of the line-profile.
Again, the radiation is after the processes of absorption and
emission redistributed over all wavelengths of the line.
Clearly, after this, radiation has more energy that is taken from
particle kinetic energy.
Consequently, the plasma is cooler now.
These processes are
%Ku1:
%Kr1: to je sice pravda, ale ja bych to tam nedaval, protoze to pro ten vyklad
%az zas tak moc dulezite neni
%amplified in the presence of the velocity gradient (due to the Doppler
%shift) and become
%
important for low-density winds \citep
%Ku1: [e.g.]
[see]
%
%Kr1: []{kkii}.
[]{go,kkii}.

\subsection{How to transfer momentum to the passive component?}
\label{trenakap}

\begin{figure}
\begin{center}
\resizebox{0.5\textwidth}{!}{\includegraphics{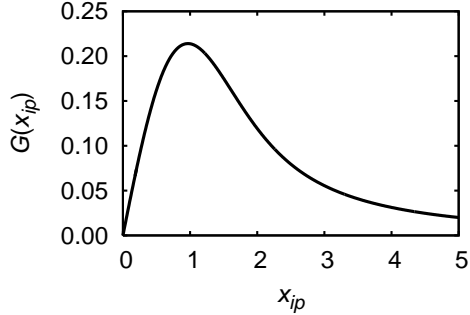}}
\end{center}
\caption{The Chandrasekhar function
%Ku1:
$G(x_{\text{ip}})$.}
\label{chandra}
\end{figure}

Since hot star winds are ionised, the most efficient way to transfer
%Kr2:
acquired
momentum from heavier elements to the passive component (H, He) is due
to the Coulomb collisions.
Frictional force on passive component (p) due to
%Ku1:
metallic
ions (i) is
%Ku1:
given by
\begin{equation}
\label{trenisil}
f_\text{pi}
%Ku1: =\rho_\text{p}g_\text{pi} - g_\text{pi} se podle mne nikde jinde
%nevyskytuje a ani neni zavedeno.
%
\approx{n}_\text{p}{n}_\text{i}\frac{4\pi
{q}_\text{p}^2{q}_\text{i}^2}{k T}\ln\Lambda
G(x_{\text{ip}}),
\end{equation}
where ${n}_\text{p}$, ${n}_\text{i}$ are number densities of wind components,
${v}_\text{i}$, ${v}_\text{p}$ are their radial velocities, and
${q}_\text{p}$, ${q}_\text{i}$ their charges.
The frictional force
%Ku1: Eq.~(\ref{trenisil}) - psat Eq. mi prijde nadbytecne
\eqref{trenisil}
depends on the velocity difference
${v}_\text{i}-{v}_\text{p}$ via the so called Chandrasekhar function
$G(x_{\text{ip}})$ (see~Fig.~\ref{chandra}), where
\begin{align}
x_\text{ip}=&\frac{|{v}_\text{i}-{v}_\text{p}|}{\alpha_\text{ip}},&
\alpha_\text{ip}^2
%=&\frac{2k\zav{m_\text{i}T_\text{p}+m_\text{p}T_\text{i}}} 
\approx\,&2kT\frac{{m_\text{i}+m_\text{p}}}
{m_\text{i}m_\text{p}}.
\end{align}
For very low velocity differences, $x_\text{ip}\lesssim0.1$, the
transfer of momentum between metallic and passive wind component is
efficient.
Wind is well-coupled in this case and it can be treated as
%Ku1: one-component.
one component.
For higher velocity differences, $x_\text{ip}\gtrsim0.1$, the frictional
heating becomes important \citep{kkii}.
For even higher velocity differences $x_\text{ip}\gtrsim1$, the
Chandrasekhar function $G(x_{\text{ip}})$ is
%Ku1:
a
decreasing
function of the velocity
%Ku1: difference, consequently
difference. Consequently,
the transfer of momentum between metallic and passive
wind
%Ku1: component
components
is inefficient and wind components may decouple \citep[e.g.][]{treni}.

\section{Passing to macro: the radiative force}

The radiative force is given by an integral
\begin{equation}
\label{mihalas}
f_\text{rad}=\frac{1}{c}\int\chi_\nu F_\nu\,\text{d}\nu.
\end{equation}
Whereas the absorption coefficient $\chi_\nu$ depends explicitly
%Kr1:
only
on local wind properties, the radiative flux $F_\nu$ 
%Ku1: obtained using the solution of the radiative transfer equation.
%Thus, the radiative flux is a non-local quantity.
is a non-local quantity and has to be obtained using the solution of the
radiative transfer equation.
Another
complication for the calculation of the radiative force arises due to the
Doppler effect. As the stellar wind is accelerated, the wavelength of a given
line at which the wind absorbs the stellar radiation shifts to shorter
wavelengths (in the reference frame of the star). 
%Kr3: This
Although this
essentially enhances the radiative force since the wind
is able to absorb stellar radiation that is unattenuated by the material below,
%Kr3: but
%
this effect also significantly complicates consistent  calculation of
the radiative force. However, in the case of the stellar wind with relatively
high velocity gradients both these effects may help to calculate the radiative
force in an
%Ku1: approximative
approximate
way. This is
%Ku1:
the
so called
%Ku1: Sobolev
\citet{sobolev}
approximation.

The radiative force due to optically thick
%Ku1: line
%Kr1: je jedna: lines
line
in the Sobolev approximation is given by
\begin{equation}
\label{siltlu}
f^\text{rad, thick}\approx\frac{\nu F_\nu}{c^2}\frac{\text{d}v}{\text{d}r},
\end{equation}
where $F_\nu$ is the stellar flux at the frequency of a given line $\nu$ and $v$
is the wind velocity. Note that the radiative force in an optically thick line
does not depend on line properties (e.g.~the occupation numbers of corresponding
levels or the line oscillator strength). This is one of the reasons why metals
that have very low number density compared to hydrogen or helium
%Ku1:
(but a large number of optically thick lines)
may be so important for the wind acceleration. 

The radiative force in
%Ku1:
an
optically thin line is given by
\begin{equation}
\label{silten}
f^\text{rad, thin}\approx\frac{\pi e^2}{m_\text{e}c^2}
F_\nu g_if_{ij}\zav{\frac{n_i}{g_i}-\frac{n_j}{g_j}}.
\end{equation}
Metals dominate also to the optically thin line force due to very
high number of their lines
%Kr3:
and due to frequent complete ionization of hydrogen and helium.

The total radiative force is given by the sum of
%Ku1: contribution Eqs.~(\ref{siltlu}),~(\ref{silten})
contributions \eqref{siltlu} and \eqref{silten}
of individual lines. The calculation can be
simplified using two approaches. First, it is possible to use the line
distribution function \citep{cak,pusle} parameterised by the set of force
multipliers $k$, $\alpha$ and $\delta$ to obtain the radiative acceleration
in CAK approximation
\begin{equation}
g^\text{rad}\sim k\,\rho^\delta_\text{el}
\zav{\frac{1}{\rho}\frac{\text{d}v}{\text{d}r}}^\alpha,
\end{equation}
where $\rho^\delta_\text{el}$ is electron density and $\text{d}v/{\text{d}r}$ is
the velocity gradient. Another line-force
parameterisation was introduced by \citet{gayley} using $\bar Q$ parameter,
\begin{equation}
g^\text{rad}\sim {\bar
Q}^{1-\alpha}\,\rho^\delta_\text{el}\zav{\frac{1}{\rho}\frac{\text{d}v}{\text{d}r}}^\alpha.
\end{equation}

\section{Macro view}

\subsection{Basic structure of 1D CAK models}

We have understood how the stellar winds of hot stars are accelerated on a
micro-level. Now we shall discuss the influence of the wind microscopic
structure on the observable macroscopic properties.

To do so, we have to calculate at least approximate stationary spherically
symmetric wind models in the CAK approximation.
However, there is a problem because an extensive table of line-force
parameters calculated by \citet{abpar} leads to 
%Kr2: overestimated
%Ku3: the
%
overestimation of
mass loss
rates
%Kr2:
\citep[e.g.][]{nltei}
%Ku1: Odkud vime, ze je mass loss overestimated? Chtelo by to citaci.
%Kr1: To jsme koneckoncu spocitali taky my, ale vsichni to vedi, takze
%bych radeji necitoval.
%Ku2: Myslim, ze vetsina ucastniku v Sapporu to nevi. Spis myslim, ze
%vsechny, kteri to vedi, osobne znas. Takze jsem pro citovani.
On the other hand, for massive stars it is possible to use
%Ku1: more realistic - Jses si uplne jisty, ze jsou 'more realistic'?
%Kr1: Urcite.
recent extended set of
force parameters of \citet{kudmet}.

It is possible to show that wind mass loss rate scales with stellar luminosity
\begin{equation}
\dot M = 4\pi\rho(r)v(r)r^2\sim L^{{1}/{\alpha'}},\quad
\alpha'=\alpha-\delta.
\end{equation}
\citep{kupul},
%Kr2:
where $\alpha$ and $\delta$ are usual CAK parameters.
This means that wind mass loss rates depend
%Kr1:
mostly
on the stellar luminosity. 
Mass loss rates  are also significantly influenced by wind properties on a
micro-level (e.g.~metallicity, ionization structure) via
%Ku1:
the %protahuje text o stranku
parameter~$\alpha'$.

Wind terminal velocity $v_\infty$ depends mostly on the
escape velocity 
\begin{equation}
v_\infty=c(T_{\mathrm{eff}})v_{\mathrm{esc}}, \qquad
c(T_{\mathrm{eff}})\approx1-3
\end{equation}
\citep[e.g.][]{lsl} and only slightly on wind properties on a micro-level.

It can be also shown that clever combination of the mass loss rate, the terminal
velocity and the stellar radius in the form of
\begin{equation}
\dot{{M}} v_\infty \zav{R_*/\text{R}_{\odot}}^{1/2},
\end{equation}
that resembles wind momentum and is therefore called the modified wind momentum
depends mostly on the stellar luminosity and only marginally on the stellar mass
%Ku1: \citep{kupul}. - k tomu by mela patrit starsi citace Kudritzkiho
%	ja si ale nepamatuji, ktera.
%Kr1: Tu jsme meli v NLTE1, ale Puls nam doporucil tudle.
\citep[e.g.][]{kupul}.

\begin{figure}[ht]
\begin{center}
\resizebox{0.50\textwidth}{!}{\includegraphics{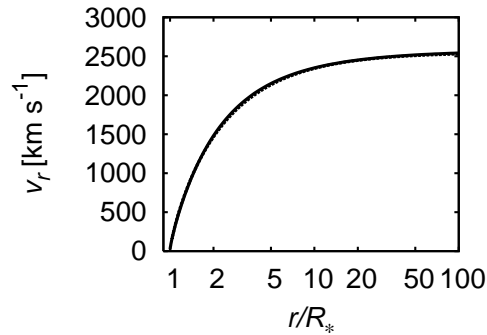}}
\end{center}
\caption{Calculated velocity structure for O6V star in the CAK approximation.
%Kr1:
The overplotted beta-velocity law (dashed line) is nearly identical.}
%Ku1: Chtelo by to nejak spojit s beta zakonem. Treba priradit parametr
%beta a namalovat zavislost, aby bylo videt, jak je beta zakon dobry ci
%spatny.
\label{rychle}
\end{figure}

%Ku1: Finally, it is possible to calculate wind models in a discussed
%approximation.
The approximation described above allows calculation of wind models.
Derived velocity structure (e.g.~Fig.~\ref{rychle}) can be in many cases
approximated by the so-called beta velocity law using parameter $\beta$
in the form 
%Kr1: dal jsem na radek
\begin{math}
v(r)=v_\infty\zav{1-
%Ku2: \frac{R_*}{r}
R_*/r
}^\beta.
\end{math}

\subsection{Multicomponent effects}

%Ku1: We have shown that - to jo, ale jindy a jinde
%
For low-density winds the transfer of momentum between
metals and hydrogen or helium may become inefficient
%Kr1:
(see Sect.~\ref{trenakap}).
There are several types
of flow with respect to the multicomponent effects \citep{kkpo}.

\subsubsection{Winds with negligible multicomponent effects}

For stars with relatively dense winds (e.g.~winds of Galactic O supergiants) the
multicomponent effects can be neglected. Such stellar wind can be adequately
treated as one-component.

\subsubsection{Wind temperature influenced by frictional heating}

For stars with lower-density winds or very low metallicity the transfer of
momentum (and energy) between metals and passive (H, He) component becomes
inefficient and part of
%Ku3: energy transferred
transferred energy
goes to heating \citep{curdi,gla}.
%Ku1: Je gla nejlepsi odkaz? Nebyl by lepsi treba kki nebo kkii?
%Kr1: Gla je nejnovejsi, tam se diskutuje i ta metalicita.

\subsubsection{Decoupling in the wind}

For stars with very low wind densities or with very low metallicities hydrogen
and helium decoupling may occur in the wind \citep{treni,kkii}. The stellar wind
is not stable in this case \citep{op,kkiii}. This problem was studied using
%Ku1: HD
%Kr1: hydrodynamic - musi se setrit mistem
%Ku2: Ach jo, HD=Henry Draper, nebo taky Hodne Dobry ...
HD
simulations \citep{reac,viktor}.

\subsubsection{Decoupling of wind components in the atmosphere}

Helium decoupling in the wind was proposed by \citet{HuGr} as the
explanation of the chemical
peculiarity of He-strong stars.
%Ku1: However, it seems that
%
Due to its low charge, helium may decouple in the stellar atmosphere
%Kr3:
of cooler stars
and a helium-free wind may exist in this case.
For extremely low-density winds also hydrogen may decouple
in the atmosphere and purely metallic stellar wind may exist
\citep{babelb}. 
%
%Ku1: Finally, 
%
For hot stars with lower luminosities
the radiative force is
%Ku1: note
not
able to expel atoms out from stellar gravitational
potential well, however the radiative force may cause chemical peculiarity
\citep[e.g.][]{mpoprad}.

\begin{figure}[ht]
\begin{center}
\includegraphics[width=0.60\textwidth]{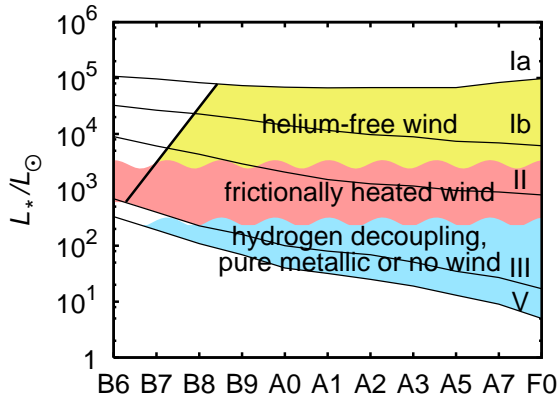}
\end{center}
\caption{Regions with different types of stellar wind in HR diagram.}
\label{hrd}
\end{figure}

\subsubsection{}
The regions in HR diagram with different types of stellar wind are given in
Fig.~\ref{hrd}. Note, however, that
NLTE models are necessary
%Kr1: tu
to
study these effects in detail. Finally,
multicomponent effects may become important
even for slightly earlier stars due to possible decoupling of individual
elements.% (Krti\v cka, submitted).

\section{Beyond classical CAK models}

\subsection{Influence of rotation}

Rotation may cause significant deviations of wind structure from the spherical
symmetry. \citet{bjoca}  proposed that discs of
%Ku1: Be
rapidly rotating
stars may be caused by the wind
compression due to the stellar rotation. However, 2D
%Ku1:
hydrodynamic
wind models
\citep{bezdisko,bezdiskp} showed that due to the gravity darkening
and nonnegligible
non-radial component of the
radiative force the wind density and mass loss rate at the equator are lower
than at the poles.
%Ku1: nasledujici mi prijde zbytecne, navic theta neni nikde definovano
% a chtelo by to proto vic vysvetlit, na coz neni misto.
%\begin{equation}
%\dot M(\theta)\sim \zav{1-\Omega^2\sin^2\theta},
%\quad\Omega=v_\text{rot}/v_\text{crit},
%\end{equation}
%where $v_\text{crit}$ is the critical rotational velocity. 

\subsection{Influence of magnetic fields}

Also magnetic fields may cause significant deviations from spherical symmetry.
They may be important for the correct explanation of the circumstellar activity of
Bp stars \citep{bamo,grohun,trojka}. According to MHD models of \citet{udo}, the
overall degree to which the wind is influenced
by the magnetic field depends on
%``wind magnetic confinement parameter" $\eta_*$ which characterises
the ratio between magnetic field energy
density and wind
kinetic energy density $\eta_*$,
\begin{equation}
\eta_*=\frac{B^2/(8\pi)}{\rho v^2/2}.
\end{equation}
For a weak confinement ($\eta_*<1$),
the magnetic field is opened by the wind outflow.
The structure of the circumstellar magnetic field is given mostly
by the stellar wind and the influence of the magnetic field on wind is not a
significant one. However, for strong confinement ($\eta_*> 1$)
the situation is essentially opposite. The flow near
the star is driven by the magnetic field. For B
stars even a moderate magnetic field intensity ($B < 100\,$G) causes
strong confinement of the circumstellar environment by the magnetic field and is
therefore very
important for its structure. 
%circumstellar environment of many B stars.
Note that for very
strong magnetic fields some parts of the circumstellar envelope may be
hydrostatic \citep{towo}. 

Magnetic fields were 
%Kr1: also invoked
used
by \citet{magdisk} to explain
%Ku1:
the
Be phenomenon.
%However, MHD models of \citet{asifsap} do not support this idea.
However, this is not supported by MHD
%Ku1: models
simulations
\citep{asifsap}.

\subsection{NLTE models}

%Ku1: "Triplet"
The "triplet"
of parameters
%Ku1: $k$, $\alpha$, $\delta$ or $\bar Q$, $\alpha$, $\delta$
($k,\alpha,\delta$) or ($\bar Q,\alpha,\delta$)
gives only rough approximation to the radiative
force. To solve this problem, it is possible to either
\begin{itemize}
%Ku1:
\itemsep -1pt
\item introduce depth-dependent radiative-force parameters
\citep[e.g.][]{pahole,kudmet}, or to
\item calculate the radiative force directly without any radiative-force
parameters \citep[e.g.]
%Ku1: []
[and references therein]
{vikolamet,nltei}.
\end{itemize}
NLTE approach is necessary in any case to obtain correct wind parameters (mainly
mass loss rates). 

The introduction of a more realistic radiative force based on the appropriate
level occupation numbers enables detailed study of
wind parameter variations.
\citet{bista} found high sensitivity of calculated wind
parameters of P~Cyg on its stellar parameters -- the {\em  bi-stability}.
\citet{vikolabis} found the bi-stability jump at around
$T_{\mathrm{eff}}\approx25\,000\,$K for normal supergiants. Using terminal
velocity measurements of \citet{lsl}
they concluded that for stars cooler than the temperature corresponding to the
bi-stability jump the mass loss rate $\dot M$ increases $5\times$,
whereas the terminal velocity $v_\infty$ decreases $2\times$.
The bi-stability jump is caused by an increase
of the line acceleration due to \ion{Fe}{iii} lines close to the stellar surface.
These calculations slightly overestimate the
correct temperature location of the
bi-stability jump, which occurs roughly at
$T_{\mathrm{eff}}\approx21\,000\,$K according to the results of
\citet{lsl}. Jump properties have still to be tested against
observations since results of \citet{trundle} show a different picture of
the jump for SMC stars \citep[c.f.][]{rychlomluvka}.

\subsection{Radiative force -- more exact approximations}

Inclusion of higher order approximations to the radiative force leads to the
wind instabilities \citep{feld}. Resulting wind shocks may (at least partly)
explain the observed X-ray emission of hot stars. Detailed discussion of this
problem can be found e.g.~in a review of \citet{owopo}.

\section{Open questions}

There are many open questions connected with the stellar wind of OB stars. Here
we discuss only those that seem to as to be especially appealing
nowadays.
%ku1:
%Kr1: to uz tam v podstate je: at least to us.
%

\subsection{Influence of instabilities and clumping}

Observed hot star wind properties show signatures of spatially organised
structures -- clumping
\citep[e.g.][]{igor,martins}. As noted above, the radiative driving is unstable
and
may cause generation of shocks \cite[e.g.][]{owopo}. The relation
between instabilities and clumping is still unclear,
%Ku1:
although clumping parameter is ofted used as an additional free
parameter of wind models.

\subsection{Winds close to $\Omega\Gamma$ limit}

During the stellar evolution some stars may come close to the $\Omega\Gamma$
limit \citep{mame}. However, it is not clear what sets the mass loss rate in
this case. The radiatively-driven wind cannot drive outflow with
%Ku1:
an
arbitrary
mass loss rate \citep{owoga}.
It is possible that to estimate these mass loss rates it is necessary
to calculate "unified" models of stellar interior and envelope. 

\subsection{What are wind mass loss rates?}

Reliable theoretical predictions of the mass loss rate are still insufficient.
One of the reasons is that detailed (NLTE) wind models are necessary to predict
mass loss rates. As the consequence, reliable predictions (for OB star
domain) are available only for
O-stars and luminous B-stars \citep[e.g.][]{pahole,vikolamet,nltei} and for hot
horizontal branch stars \citep{vicas}. Note that even frequently used predictions of
\citet{vikolamet} suffer from many 
%Kr2: approximations
%Ku4: questionable - radil jsem se s Adelou, questionable je prilis
%silne slovo a s Vinkem jsem byl v hospode
%
simplifications
like neglect of "line
branching" \citep{sim}, wind instabilities and clumping or X-ray radiation \citep{macown}. 
Whereas there is a relatively good agreement between theoretically predicted
mass loss rates and mass loss rates derived from observations for individual
hot OB stars, there is a significant disagreement between these values
for cooler B supergiants \citep{vikola}. Although part of this discrepancy may
be due to observations, some part of this discrepancy is likely due to model
simplifications.

Even worse, for many stars (e.g.~for many main-sequence B stars) there
are no reliable predictions of mass loss rate available. The problem is that
there are several processes that may influence the mass loss rate of these
stars, e.g.
\begin{itemize}
%Ku1:
\itemsep -1pt
\item multicomponent wind structure \citep{treni,curdi,babela,kkii},
\item GO (Doppler) heating \citep{go},
\item thin-wind 
%Kr3: instability close to the sonic point
effect
\citep{owpu}.
\end{itemize}

Low-density stellar winds are also very difficult to observe. However, some
indirect wind indications may be available, e.g.~due to certain chemical
peculiarities \citep{nedolez,dwobune} or due to wind magnetic braking of stellar
rotation \citep{miktady,oksala}.

\begin{figure}[ht]
\begin{center}
\includegraphics[width=0.60\textwidth]{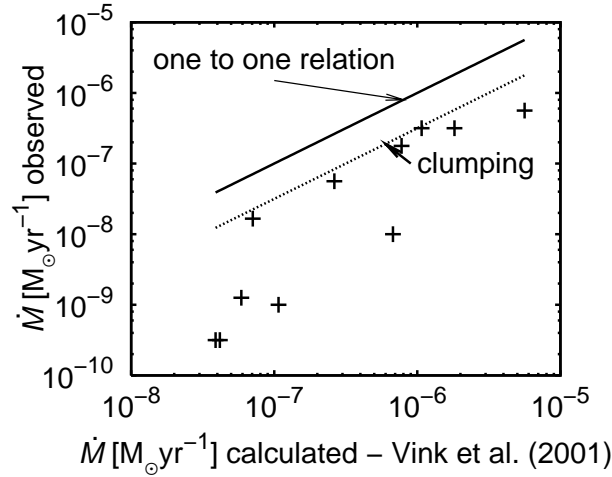}
\end{center}
\caption{Comparison of mass loss rate derived from observations allowing for
wind clumping and the theoretical (smooth) mass loss rates}
\label{martins}
\end{figure}

There is an increasing observational evidence that hot star winds are clumped.
\citet{martins} 
%Kr3: derived mass loss rates for
studied winds of
several Galactic O-stars
%Kr3:
in detail
and concluded that 
%Kr3: their
%
mass loss rates
%Kr3:
derived assuming clumped winds
may be $2\times-\;5\times$ lower than 
%Kr3: predicted smooth ones due to the clumping
%Ku4: that
those
derived assuming smooth winds
(Fig.~\ref{martins}). 
Finally, for some stars with low luminosities the wind mass loss rates seem
to be more than
ten times lower than the predicted ones
%Kr3: 
(Fig.~\ref{martins}).
Similar effect was detected for SMC
stars by \citet{bourak}.
Thin-wind
%Kr3: instability
effect
%Kr1:
\citep{owpu}
may 
%Kr4: cause
help to understand the origin of
this difference.
%Kr1: \citep{owpu}.

\section{Conclusions}

We have discussed
%Ku3: the
%
basic physical principles that drive the stellar winds of
%Kr1: hot
OB
stars both on micro and on macro
%Ku1: level.
levels.
We have also seen that despite the
tremendous effort of many astronomers, more than 30~years after the publication
of
%Ku1:
the
seminal CAK paper we are still not sure what
%Ku1: are the mass loss rates of hot stars.
%  tohle mi systematicky opravovala Katrien :-)
the mass loss rates of hot stars are.
To conclude with something
%Ku3: that is
%
more optimistic we switch to slightly
different stars -- WR stars.
\cite{graham} were able to theoretically explain the acceleration of
the stellar wind of WR stars.
According to these models, wind sonic point is located deep in the
stellar interior and the stellar wind in the inner regions is
accelerated due to the iron opacity bump
%Ku1: Neni iron opacity bump jako iron opacity bump. Jses si jisty, ze
%ten bump je zpusobem prechodem mezi stejnymi ionizacnimi stupni ve
%vsech pripadech?
%Kr1: Konvekci zminuji uz Graefener a Hamann; co se tyce pulzaci, pak ty maji
%svou pricinu take diky skoku v opacite okolo 200 000 K (apspon podle clanku).
%Ku2: Neodpovedel jsi na otazku, ptal jsem se na stejne ionizacni stupne.
%Kr2: Tech ionizacnich stupnu je vic, ale odpovidaji si vzajemne.
%Ku4: tak to svedeme na ne
%Note, however, that
and
convection in WR stars occurs also as a consequence
of the iron opacity bump.
%Ku3: tady bychom meli pridat odkaz, ja ho nevim
%Kr3: Ja taky ne, Graefener \& Hamann nic neuvadi.
%Ku3: and pulsation in $\beta$~Cep variables are due to the iron opacity
%bump \citep{moska,cox}.
Iron opacity bump stands also behind pulsations in $\beta$~Cep
stars \citep{moska,cox}.
Consequently,
%Ku3: the same physical process causes
processes of the same physical nature cause
completely different behaviour in
different
%Ku3: circumstances! - nerad kricim ...
%Kr3: nekricime, vyjadrujeme podiv :-) circumstances.
%Ku4: Tak to bychom meli vymyslet novy interpunkcni znak: divnik :-)
circumstances!
Since wind critical point in the case
%Ku1: ow
of
WR stars is located relatively deep in the stellar interior, any
differentiation between stellar core, atmosphere, and wind may become
artificial and we might need to (at least for some
specific problems) calculate "unified" models of stellar interior
and exterior.
%Kr1:
%Apparently, the only possibility how to understand 
%hot stars' winds
%in detail from the theoretical point of view is using
%detailed modelling.

\acknowledgements We would like to thank to our colleagues from our home
institutes for
%Ku1: the discussion of
discussing
this topic. This work was supported by grants GA
\v{C}R 205/03/D020, 205/04/1267.
The Astronomical Institute Ond\v{r}ejov is supported by project Z1003909.

%%%%%%%%%%%%%%%%%%%%%%%%%%%%%%%%%%%%%%%%%%%%%%%%%%%

\newcommand\krtek{Krti\v cka}

%Ku1: Owocki:
\noindent {\bfseries Owocki:}
I must disagree with the notion that Fe opacity bump can be an important
mechanism for wind driving. I think to focus on using this bump to set the wind
through the sonic point is mis-guided. The sound speed is of order
$25\,\text{km}\,\text{s}^{-1}$, about 30~times smaller than the escape speed
$v_\text{esc}$ ($\approx 700\,\text{km}\,\text{s}^{-1}$); implying an energy
that is $1/1000$ of that needed to escape the star Thus even if Fe opacity sets
the wind through the sonic point, we must appeal to some other mechanism to do
the $99.9\%$ of the remaining work needed to escape. I think understanding wind
mass loss should focus instead on the mechanism that does this overwhelming bulk
of the work.

%Ku1: \krtek:
\noindent {\bfseries \krtek:}
I agree that iron bump opacity alone is not able to drive the stellar
wind of WR star and does only a certain part of work necessary to launch the wind. Also
other opacity sources
%Ku1: shall
should
be included. To drive
%Ku1:
the
wind it is necessary to
accelerate the wind material from the subsonic velocities to the velocity higher
than the escape speed. 

%Ku1: Townsend:
\noindent {\bfseries Townsend:}
Another way of thinking about the iron opacity bump is to recall that
it drives SPB \& $\beta$ Cep pulsations. The reason these stars show pulsations
is that the iron opacity bump disappears when the temperature varies
significantly from $200\,000\,$K. This makes it very difficult to understand how
a wind -- which requires a spatially-extended region of high opacity -- could be
driven by the iron opacity bump.

%Ku1: \krtek:
\noindent {\bfseries \krtek:}
\cite{graham} were able to consistently accelerate the wind from the
subsonic velocities to the velocities higher than the escape speed.
%Kr1: The pulsational argument is not valid in this case (in my view)
I do not think that your argument is important,
because the physical
mechanism that drives the pulsations is different. Pulsations are
%Kr1: , to my knowledge,
driven basically by the heat flux, however the stellar wind is driven
by the gradient of the radiation pressure.

%Ku1: Stee:
\noindent {\bfseries St\'ee:}
The terminal velocity and the mass flux is strongly depending on the
inclination angle (on the stellar colatitude) i.e.~larger at the poles and lower
at the equator. Thus it is normal that there is a large scattering in the
mass loss rate versus effective temperature graph. Moreover the mass loss also
%Ku1: depend - ted nevim, maji se opravovat chyby v anglictine u
%tazatelu?
depends
on the lines you are using to determine it (IR, visible, UV). Finally, it
seems difficult for me to compare observational and theoretical $v_\infty$ and
$\dot M$ for nonspherical wind (for instance following the bi-stability scheme
you have shown).

%Ku1: \krtek:
\noindent {\bfseries \krtek:}
%
%Kr1: Stellar 
Rotation is 
%Kr1: indeed
important for wind structure, however
%Ku1:
it
has only second order effect for stars with rotational velocities well
bellow the critical
%Kr1: rotational velocity
one
\citep{bezdiskp}.
It is not likely that it causes
%Kr1: very high (order of magnitude)
order of magnitude
differences of 
%Kr1: mass loss rates
$\dot M$ 
for cool B stars or
too high predicted  
%Kr1: mass loss rate
$\dot M$ 
for low-luminosity stars.

\end{document}